\documentclass[useAMS,usenatbib]{mn2e}
\usepackage{rotating}
\usepackage{journals}

\def\approxgt{\ifmmode \rlap{$>$}{}_{{}_{{}_{\textstyle\sim}}} \else%
$\rlap{$>$}{}_{{}_{{}_{\textstyle\sim}}}$\fi} 
\def\approxlt{\ifmmode \rlap{$<$}{}_{{}_{{}_{\textstyle\sim}}} \else%
$\rlap{$<$}{}_{{}_{{}_{\textstyle\sim}}}$\fi}

\def\farcs{\hbox{$.\!\!^{\prime\prime}$}}

\def\arcsec{\hbox{$^{\prime\prime}$}}

\def\flx{erg cm$^{-2}$ s$^{-1}$}
\def\lum{erg s$^{-1}$}

\def\chan{{\it Chandra}}
\def\src{H~1743-322}
\def\swift{{\it Swift}}
\newcommand{\ledd}{$L_{Edd}$}

\newcommand{\mdot}{$\dot{m}$}

\newcommand{\lx}{$L_X$}

\LARGE \normalsize \title[The outburst decay of \src]{Following the 2008 outburst decay of the black hole candidate \src\, in X--ray and radio}

\author[Jonker et al.]  {P.G.~Jonker$^{1,2}$\thanks{email : p.jonker@sron.nl},
J.~Miller--Jones$^3$, J.~Homan$^{4}$, E.~Gallo$^{4}$, M.~Rupen$^{5}$, J.~Tomsick$^{6}$,\newauthor R.P.~Fender$^{7}$, P.~Kaaret$^{8}$, D.T.H.~Steeghs$^{9,2}$, M.A.P.~Torres$^{2}$, R.~Wijnands$^{10}$, \newauthor
S.~Markoff$^{10}$, W.H.G.~Lewin$^{4}$\\ 
$^1$SRON, Netherlands Institute for Space Research,
Sorbonnelaan 2, 3584~CA, Utrecht, The Netherlands\\ 
$^2$Harvard--Smithsonian Center for Astrophysics, 60 Garden Street, Cambridge, MA~02138, U.S.A.\\
$^3$NRAO Headquarters, 520 Edgemont Road, Charlottesville, VA 22903, USA\\
$^4$MIT, Kavli Institute for Astrophysics and Space Research, 70 Vassar Street, Cambridge, MA 02139, USA\\ 
$^5$NRAO, Array Operations Center, 1003 Lopezville Road, Socorro, NM 87801, USA\\
$^6$Space Sciences Laboratory, University of California, Berkeley, USA\\
$^7$School of Physics and Astronomy, University of Southampton, Southampton SO17 1BJ\\
$^8$Department of Physics and Astronomy, University of Iowa, Van Allen Hall, Iowa City, IA 52242, USA\\
$^9$Department of Physics, University of Warwick, Coventry CV4 7AL\\
$^{10}$Astronomical Institute `Anton Pannekoek', University of Amsterdam, Postbus 94249, 1090 GE Amsterdam, the Netherlands\\
}
\begin{document}

\maketitle

\begin{abstract} \noindent In this Paper we report on radio (VLA and
  ATCA) and X--ray ({\it RXTE}, \chan\ and \swift) observations of the
  outburst decay of the transient black hole candidate \src\, in early
  2008.  We find that the X--ray light curve followed an exponential
  decay, levelling off towards its quiescent level. The exponential
  decay timescale is $\approx$4 days and the quiescent flux
  corresponds to a luminosity of 3$\times 10^{32}(\frac{{\rm d}}{7.5
    {\rm kpc}})^2$\lum. This together with the relation between
  quiescent X--ray luminosity and orbital period reported in the
  literature suggests that \src\ has an orbital period longer than
  $\approx$10 hours. Both the radio and X--ray light curve show
  evidence for flares. The radio -- X--ray correlation can be well
  described by a power--law with index $\approx$0.18. This is much
  lower than the index of $\approx$0.6--0.7 found for the decay of
  several black hole transients before. The radio spectral index
  measured during one of the radio flares while the source is in the
  low--hard state, is -0.5$\pm$0.15, which indicates that the radio
  emission is optically thin. This is unlike what has been found
  before in black hole sources in the low--hard state. We attribute
  the radio flares and the low index for the radio -- X--ray
  correlation to the presence of shocks downstream the jet flow,
  triggered by ejection events earlier in the outburst. We find no
  evidence for a change in X--ray power law spectral index during the
  decay, although the relatively high extinction of ${\rm N_H \approx
    2.3 \times 10^{22} cm^{-2}}$ limits the detected number of soft
  photons and thus the accuracy of the spectral fits.

\end{abstract}

\begin{keywords} stars: individual (\src) --- 
accretion: accretion discs --- stars: binaries 
--- X-rays: binaries
\end{keywords}

\section{Introduction} 

A wide variety of astrophysical objects are thought to be powered by
accretion.  In the case of accretion onto a black hole, the properties
of these flows mainly depend on the mass and spin of the black hole
and on the mass accretion rate (\mdot). Variations in \mdot\ are
probably responsible for the spectral and variability states of black
hole X-ray binaries (BHXBs; \citealt{2006csxs.book..157M}).

Transient BHXBs spend long periods at very low X-ray luminosities,
referred to as `quiescence', but during occasional outbursts the
luminosity increases by as much as 7 to 8 orders of magnitude,
typically reaching values of tens of per cent of the Eddington
luminosity (\ledd). \citet{2003A&A...409..697M} has shown that below
around a few percent of \ledd\ BHXBs typically return to the
so--called hard state, in which the spectrum is dominated by a power
law component with an index $\sim$1.5
(e.g.~\citealt{2005ApJ...622..508K}).  BHXBs are still in that state
when they have decayed down to $10^{-4}$\,\ledd.

With {\it Chandra} and {\it XMM-Newton} it is possible to explore the
spectral properties of BHXBs all the way to quiescence, for which
log(\lx) is typically $\sim$30.5--33.0 (e.g.\
\citealt{2002ApJ...570..277K}; \citealt{2003A&A...399..631H}).  This
range corresponds to $\sim$3$\times 10^{-9}-$9$\times 10^{-7}$\,\ledd\
for a 8.5$M_\odot$ black hole (which is the average black-hole mass
from \citealt{2006csxs.book..157M}). About 15 BHXBs have been observed
in quiescence with {\it Chandra} and {\it XMM-Newton}, but only in a
handful of cases could the spectral properties be well constrained.
Nevertheless, these few cases suggest an interesting property of
quiescent spectra: when fitted with a power-law the indices tend to be
considerably softer than the index of $\sim1.5$ found in the hard
state (e.g.~\citealt{2006ApJ...636..971C}). Although the errors for
individual sources are still large, the overall trend seems to
indicate that, at least spectrally, the quiescent state is different
from the hard state that is observed three to five decades higher in
\lx.

Corbel et al.~(2003) and \citet{2003MNRAS.344...60G} have demonstrated
the existence of a relation between observed X-ray and radio emission
from hard state black holes, of the form $L_{\rm radio} \propto L_{\rm
  X}^{0.7}$; this relation is nearly the same for V404~Cyg and
GX~339--4. However, \citet{2007A&A...466.1053X},
\citet{2007ApJ...659..549C}, \citet{2007ApJ...655L..97R},
\citet{2007ApJ...655..434S} and \citet{2008MNRAS.389.1697C} found that
the relation might not be as universal as previously thought (see the
compilation in \citealt{2007AIPC..924..715G}). On the other hand
\citet{2005ApJ...624..295H} and \citet{2006MNRAS.371.1334R} found
further relations between the X--ray and near--infrared fluxes in
black hole transients: $L_{IR} \propto L_X^{\,\sim0.6}$. Both
relations extend over more than three decades in luminosity down to
$\sim 10^{-4}$\,\ledd\ and suggest that the X--ray, near--infrared and
radio emission in the hard state are intimately connected.
Correlations between optical and X--ray light in the black hole source
GX~339--4 point in the same direction (\citealt{2008MNRAS.390L..29G}).
{\it Chandra} and VLA observations of A~0620--00 in quiescence by
\citet{2006MNRAS.370.1351G} showed that the radio and X--ray flux lie
on the extension of the $L_{\rm radio} \propto L_{\rm X}^{0.7}$
correlation, suggesting that it holds all the way down to quiescence.
Some models predict that the relation should break down
(i.e.~significantly steepen) around $10^{-5}$\,\ledd\
(\citealt{2005ApJ...629..408Y}). The result of Gallo et al.~(2006)
seems to rule this out, but given the contradicting results on the
decay rate and normalisation of the radio -- X--ray correlations, this
needs to be investigated further.

ADAF and jet--models both predict that the X--ray luminosity scales as
$\dot{m}^2$ below $10^{-4}$\,\ledd\ (Narayan et al.~1997;
\citealt{2003MNRAS.343L..99F}). Observational evidence for this
scaling was deduced from X-ray/radio correlations
(\citealt{2003A&A...397..645M}; \citealt{2006MNRAS.369.1451K}). In
ADAF models a gradual softening of the power--law photon index is
expected until an index of $\sim$2.2 in quiescence
(\citealt{esmcna1997}; \citealt{1997ApJ...482..448N}). However, in
ADAF models the disk should recede in the hard state, something that
is recently debated by \citet{2006ApJ...652L.113M,2006ApJ...653..525M}
(but see \citealt{2008MNRAS.388..753G};
\citealt{2009MNRAS.394.2080H}).

\src\, was discovered in August, 1977 by the Ariel--5 all--sky
monitor. The position was more accurately determined by HEAO--I ruling
out an association with 4U~1755--388 (\citealt{1977IAUC.3099S...1K};
\citealt{1977IAUC.3106....4K}). The source has shown repeated
outbursts since 2003 after being rediscovered by the International
Gamma--ray Astrophysics Laboratory (INTEGRAL; IGR~J17464-3213
\citealt{2003ATel..132....1R}).  \citet{2003ATel..133....1M} suggested
that the INTEGRAL--found source is the same as \src. Recently, the
source showed several outbursts; one in 2007/2008 (the data presented
in this paper is from this outburst;
\citealt{2008ATel.1348....1K}), one in late 2008
(\citealt{2009A&A...494L..21P}) and one in 2009
(\citealt{2009ATel.2058....1K}).

\citet{2006csxs.book..157M} list \src\ as a strong black hole
candidate.  Phase resolved optical or near--infrared observations to
determine the mass function of the black hole have not yet been
reported, even though the source was found in outburst in
near--infrared and optical bands (\citealt{2003IAUC.8112....2B};
\citealt{2003ATel..146....1S}).  Near--infrared observations of the
source in quiescence show an unrelated star within 1\arcsec\, and
found the counterpart $K$--band magnitude to be 17.1
(\citealt{2009ApJ...698.1398M}). \src\, is one of the few sources
where X--ray jets have been imaged
(\citealt{2005ApJ...632..504C}). The other sources where an X--ray jet
has been found are the black hole sources 4U~1755--33 and
XTE~J1550--564 (\citealt{2003ApJ...586L..71A};
\citealt{2002Sci...298..196C}), the neutron star source Cir~X--1
(\citealt{2007ApJ...663L..93H}; \citealt{2009MNRAS.397L...1S}) and the
peculiar source SS~433 (\citealt{2002Sci...297.1673M}). In addition to
X--ray jet outflows \citet{2006ApJ...646..394M} found a strongly
variable disc wind from \src. The distance to this source is so far
not well constrained, however, \citet{2005ApJ...632..504C} find from
the fact that the jet speed is limited to maximally the speed of light
and the observed X--ray jet proper motion that the upper limit to the
distance is 10.4$\pm$2.9 kpc, consistent with the often assumed
Galactic Center distance of 7.5 or 8 kpc. This assumption is based on
the Galactic coordinates of the source ($l=$357.255 and $b=$-1.83); it
lies towards the Galactic bulge. In this paper we assume a
distance of 7.5 kpc.

Here, we report on contemporary {\it RXTE}, \swift\ and \chan\ X--ray
and Very Large Array (VLA) radio observations of \src\, aimed at
following the X--ray and radio light curves and establishing the X--ray
-- radio correlation while the source decays to quiescence. The
observations have been obtained in the last stages of the outburst
ending early 2008.  For this outburst \citet{2008ATel.1378....1K}
reported that the source returned to the low--hard state between Jan,
24, 2008 and Feb., 1, 2008. To determine the quiescent spectral
properties we further analysed three \chan\, observations of \src\,
obtained when it was in quiescence (\citealt{2006ApJ...636..971C}).

\section{Observations, analysis and results} 

\subsection{{\it Chandra} X--ray observations} 

We observed \src\, with the \chan\, satellite using the
back--illuminated S3 CCD--chip of the Advanced CCD Imaging
Spectrometer (ACIS) detector (\citealt{1997AAS...190.3404G}) on
several occasions during the final parts of the decay to quiescence.
During the first two observations in 2008 (Obs IDs 8985 and 8986) we
employed the High Energy Transmission Grating (HETG) to mitigate
effects of photon pile--up. In the subsequent observations we windowed
the ACIS--S CCD, providing a frame time of 0.4104~s. We have
reprocessed and analysed the data using the {\it CIAO 4.0.2} software
developed by the Chandra X--ray Center. In our analysis we have
selected events only if their energy falls in the 0.3--7 keV range.
All data has been used, as background flaring is very weak or absent in
all data.

\begin{table*}
\caption{A journal of the \chan\ observations.}
\label{log}
\begin{center}
\begin{tabular}{ccccc}
\hline
Obs ID & Observing  & MJD & Time on & Count rate\\
& date & (UTC) & source (ks) & 0.3-7 keV (cnt s$^{-1}$)\\
\hline
4565$^a$ & 2004 Feb.~13 & 53047.84959 & 23.0 & $(1.5\pm0.3)\times10^{-3}$\\
4566$^a$ & 2004 Mar.~25 & 53088.74413 & 28.4 & $(0.7\pm0.2)\times10^{-3}$\\
4567$^a$ & 2004 Mar.~27 & 53091.25914 & 40.0 & $(1.8\pm0.2)\times10^{-3}$\\
8985$^{b,d}$ & 2008 Feb.~19 & 54515.14003 & 6.4  & $0.167\pm0.005$$^c$\\
8986$^{b,d}$ & 2008 Feb.~25 & 54521.04074 & 7.6  & $(8.4\pm0.3)\times10^{-2}$$^c$\\
8987$^d$ & 2008 Mar.~02 & 54527.13111 & 6.5	& 0.101$\pm$0.004\\
8988$^d$ & 2008 Mar.~08 & 54533.69200 & 13.7	& $(1.9\pm0.1)\times10^{-2}$\\
8989$^d$ & 2008 Mar.~16 & 54541.23027 & 20.5	& $(4.0\pm0.4)\times10^{-3}$\\
9833     & 2008 Mar.~17 & 54542.08033 & 11.0	 & $(2.5\pm0.5)\times10^{-3}$\\
9838$^d$ & 2008 Mar.~21 & 54546.42945 & 23.8	 &$(1.8\pm0.3)\times10^{-3}$\\
8990     & 2008 Mar.~22 & 54547.32918 & 21.2	 &$(1.6\pm0.3)\times10^{-3}$\\
9839     & 2008 Mar.~23 & 54548.30002 & 28.7	 &$(1.4\pm0.2)\times10^{-3}$\\
9837     & 2008 Mar.~24 & 54549.22579 & 20.6	  &$(1.7\pm0.3)\times10^{-3}$\\
\end{tabular}		      
\end{center}
{\footnotesize $^a$ Observations reanalysed from \citealt{2006ApJ...636..971C}. \\
 $^b$ High energy transmision grating inserted\\
 $^c$ Zeroth order count rate.\\}
{\footnotesize$^d$ Data point used in Figure ~\ref{rxcorr}.}\\
\end{table*}

Using {\sl wavdetect} we detect \src\ in each of the observations. We
have selected a circular region of 10 pixel ($\approx$5\arcsec) radius
centered on the accurately known source position (Steeghs et al.~2003)
to extract the source counts. Similarly, we have used a circular
annulus with inner and outer radius of 20 and 40 pixels centered on
the source position to extract background counts. The source and
background region for observation 4566 and 4567 excludes the X--ray
jets found by \citet{2005ApJ...632..504C}. We have made redistribution
and auxilliary response matrices for the source region of each of the
observations separately.

The net, background subtracted, source count rate for each observation
is given in Table~\ref{log}. Using {\sl xspec} version 11.3.2p
(\citealt{ar1996}) we have fitted the spectra of \src\ using $\chi^2$
statistics requiring at least 10 counts per spectral bin for the
observations with IDs 8985, 8986, 8987, and 8988 and Cash statistics
(\citealt{1979ApJ...228..939C}) modified to account for the
subtraction of background counts, the so called
W--statistics\footnote{see
  http://heasarc.gsfc.nasa.gov/docs/xanadu/xspec/manual/} for the
other observations. We have used an absorbed power--law model ({\sl
  pegpwrlw} in {\sl xspec}) to describe the data. For the observations
with IDs 8985, 8986, 8987, and 8988 we allow all three parameters
(neutral hydrogen column density ${\rm N_H}$, the power--law index and
normalisation) to float freely. Due to the relatively low number of
counts we fix the interstellar extinction during the fits to the rest
of the observations to 2.3$\times 10^{22}$ cm$^{-2}$, consistent with
the ${\rm N_H}$ found by \citet{2006ApJ...646..394M} and for the
observations with IDs 8985, 8986, 8987, and 8988. The power law index
and normalisation were allowed to float.

We list the results of our spectral analysis in Table~\ref{spec}.
Given the similar fluxes and contemporaneous observations we have also
analysed the spectral behaviour of \src\ for the observations 9838,
8990, 9839, 9837 together. Furthermore, in order to determine the
best--fitting power law index for the source in quiescence we have
fitted the spectral model to observations 4565, 4566, 4567, 9838,
8990, 9839, 9837 together (see Table~\ref{spec}).

\begin{table*}

\caption{Best fit parameters of the spectra of \src. PL refers to power law.  All quoted errors are at the 68 per
cent confidence level.}

\label{spec}
\begin{center}
\begin{tabular}{cccccc}
\hline
Obs ID &  N$_H~\times10^{22}$ & PL ~flux (0.3-7 keV \chan) & PL index & Unabs. 0.5--10 keV flux & Goodness / ${\rm \chi^2_{red}}$ \\ 
&  cm$^{-2}$ & (0.5-10 keV \swift) $\times10^{-12}$ erg$^{-1}$~cm$^{-2}$~s$^{-1}$  & &  (erg cm$^{-2}$ s$^{-1}$) & per cent / d.o.f.\\
\hline
\hline
8985$^d$ &  1.8$\pm$0.4 & 11.8$\pm$1.0 & 0.87$\pm$0.18 & $(1.8 \pm 0.2)\times 10^{-11}$  & 1.12/89\\
8986$^d$ &  1.7$\pm$0.6 &  6.3$\pm$1.0 & 1.03$\pm$0.27 & $(9^{+2}_{-0.3})\times 10^{-12}$ & 0.81/55 \\
8987 &  2.2$\pm$0.3 &  2.8$\pm$0.5 & 1.68$\pm$0.21 & $(3^{+0.6}_{-0.1})\times 10^{-12}$  & 0.86/56\\
8988 &  2.2$\pm$0.7 &  0.55$^{+0.34}_{-0.15}$& 1.75$\pm$0.47 & $(5^{+3}_{-0.5})\times 10^{-13}$  & 0.7/20\\
8989 &  2.3$^a$	  & $(1.6\pm0.3)\times 10^{-1}$ & 2.00$\pm$0.25 & $(1.5\pm0.2)\times 10^{-13}$ & 52\%\\
9833 &  2.3$^a$   & $(7.6^{+2.7}_{-1.7})\times 10^{-2}$ & 1.7$\pm$0.5 & $(8.4\pm1.7)\times 10^{-14}$ & 54\%\\
9838 &  2.3$^a$   & $(6.1\pm1.5)\times 10^{-2}$ & 1.8$\pm$0.4 & $(6.2\pm1.0)\times 10^{-14}$ & 58\%\\
8990 &  2.3$^a$   & $(6.2^{+2.9}_{-1.6})\times 10^{-2}$ & 2.1$\pm$0.4 & $(5.6\pm1.0)\times 10^{-14}$ & 76\%\\
9839 &  2.3$^a$   & $(3.9\pm0.7)\times 10^{-2}$ & 1.3$\pm$0.4 & $(4.9\pm0.7)\times 10^{-14}$ & 42\%\\
9837 &  2.3$^a$	  & $(9\pm4)\times 10^{-2}$ & 2.5$\pm$0.4 & $(7\pm2)\times 10^{-14}$ & 30\%\\
4565 &  2.3$^a$   & $(4.9\pm1.3)\times 10^{-2}$ & 1.8$\pm$0.4 & $(5\pm1)\times 10^{-14}$ & 65\%\\
4566 &  2.3$^a$   & $(2.5^{+1.3}_{-0.7})\times 10^{-2}$ & 1.8$\pm$0.6 & $(2.6\pm0.6)\times 10^{-14}$ & 71\%\\
4567 &  2.3$^a$   & $(5.9\pm1.1)\times 10^{-2}$ & 1.8$\pm$0.3 & $(6.2\pm0.7)\times 10^{-14}$ & 95\%\\
Comb$^b$ & 2.3$^a$ &  $(5.1\pm0.5)\times 10^{-2}$  & $1.84\pm0.14$ & $(5.2\pm0.3)\times 10^{-14}$ & 85\%\\
Comb$^c$ & 2.3$^a$ &  $(5.6\pm0.8)\times 10^{-2}$  & $1.90\pm0.19$ & $(5.6\pm0.4)\times 10^{-14}$ & 57\%\\
\hline
00031121001 & 2.3$^a$  & 293$\pm$6 & 1.74$\pm$0.05 & $(2.93\pm0.06)\times 10^{-10}$ & 0.97/270 \\
00031121002 & 2.3$^a$  & 106$\pm$3 & 1.59$\pm$0.08 & $(1.06\pm0.03)\times 10^{-10}$ & 1.01/166 \\
00031121003 & 2.3$^a$  & 45$\pm$3 & 1.7$\pm$0.08 & $(3.8\pm0.1)\times 10^{-11}$ & 0.67/91 \\
00031121004 & 2.3$^a$  & 24$\pm$1 & 1.6$\pm$0.1 & $(2.1\pm0.1)\times 10^{-11}$ & 0.95/73 \\
00031121005 & 2.3$^a$  & 10$\pm$1 & 1.9$\pm$0.2 & $(7.6\pm0.6)\times 10^{-12}$ & 0.63/19 \\
\hline
00031121001 & 1.85$\pm$0.13  & 260$\pm$10 & 1.51$\pm$0.08 & $(2.6^{+0.2}_{-0.03})\times 10^{-10}$ & 0.94/269 \\
00031121002 & 1.26$\pm$0.17  & 90$\pm$3 & 1.09$\pm$0.12 & $(8.9^{+0.7}_{-0.01})\times 10^{-11}$ & 0.88/165 \\
00031121003 & 2.41$\pm$0.28  & 46$\pm$5 & 1.77$\pm$0.15 & $(3.9^{+0.4}_{-0.01})\times 10^{-11}$ & 0.68/90 \\
00031121004 & 2.06$\pm$0.35  & 23$\pm$2 & 1.47$\pm$0.18 & $(2.1^{+0.2}_{-0.02})\times 10^{-11}$ & 0.95/72 \\
00031121005 & 1.7$\pm$0.7    & 9$\pm$2 & 1.6$\pm$0.4 & $(7.1^{+2.0}_{-0.1})\times 10^{-12}$ & 0.63/18 \\
\end{tabular}

{\footnotesize$^a$ ${\rm N_H}$ has been fixed to 2.3$\times10^{22}$ cm$^{-2}$.}\\
{\footnotesize$^b$ Fit to Obs ID 4565, 4566, 4567, 9838, 8990, 9839, 9837 combined.}\\
{\footnotesize$^c$ Fit to Obs ID 9838, 8990, 9839, 9837 combined.}\\
{\footnotesize$^d$ Fit parameters affected by pile--up. }\\
\end{center}
\end{table*}

\subsection{{\it Swift} X--ray observations}

The {\it Swift} satellite observed the field of \src\ during the decay
of its 2008 outburst. We here report on results obtained using the
X--ray telescope (XRT). A log of the observations is presented in
Table~\ref{swiftlog}. The {\sl ftools} software package tool {\sl
  xselect} has been used to extract source and background photons from
regions centered on the known source position or a source free region
on the CCD, respectively. Square boxes were used as extraction regions
in the first two observations (Obs IDs 00031121001 and 00031121002)
that were obtained in windowed timing mode. In that mode ten rows are
read--out compressed into one, and the central 200 columns of the CCD
are read.  Hence, there is only one-dimensional spatial information
allowing for faster CCD read--out and reducing effects of photon
pile--up. The square boxes have a size of 40 pixels ($\approx
94\arcsec$) along the one available spatial dimension.  Events were
selected if their event--grade is
0--2\footnote{http://swift.gsfc.nasa.gov/docs/swift/analysis/xrt\_swguide\_v1\_2.pdf}.
For the four remaining observations two--dimensional information is
available as the CCD is read--out in frame--transfer mode. These
observations are in what is called photon--counting mode. Circular
extraction regions that have a radius of 50 arcseconds were
used. Events were selected if their event--grade is 0--4.

We have determined spectral parameters in the same way as we did for
the {\it Chandra} observations restricting the fits to photon energies
in the range from 0.5--10 keV. We have used two approaches. In the
first we keep the neutral hydrogen column density fixed to 2.3$\times
10^{22}$ cm$^{-2}$. All the spectral fits to the {\it Swift} data were
done using a $\chi^2$ minimisation technique with 10 or more counts
per spectral bin, except the spectral fits to the last observation
(Obs.~ID 00031121006). There, we employed the Cash statistics for
estimating the best--fitting parameters, however, due to the low
number of counts we could not extract meaningful spectral parameters.
In the second approach we leave the neutral hydrogen column density
free. The results of the spectral fits using both approaches are
presented in Table~\ref{spec}.

\begin{table}
\caption{A journal of the {\it Swift} XRT observations.}
\label{swiftlog}
\begin{center}
\begin{tabular}{lccc}
\hline
Obs ID & Observing  & MJD & Time on \\
& date & (UTC) & source (ks) \\
\hline
00031121001$^a$ & 2008 Feb.~12 & 54508.06499 & 1.74\\
00031121002$^a$ & 2008 Feb.~15 & 54511.61268 & 2.26\\
00031121003$^{b,c}$ & 2008 Feb~.19 & 54515.02349 & 3.35\\
00031121004$^b$ & 2008 Feb.~22 & 54518.16801 & 4.62\\
00031121005$^{b,c}$ & 2008 Feb.~26 & 54522.05481 & 3.38\\
00031121006$^b$ & 2008 Mar.~04 & 54529.55583 & 3.53\\
\end{tabular}
\end{center}
{\footnotesize$^a$ Windowed timing observing mode.}\\
{\footnotesize$^b$ Photon counting observing mode.}\\
{\footnotesize$^c$ Data point used in Figure ~\ref{rxcorr}.}
\end{table}

\subsection{{\it RXTE} X--ray observations }

In addition to the \swift\ and \chan\ observations we have used
several archival RXTE observations in order to follow the flux
evolution of the source after the source entered the low--hard state
on the way to quiescence. We do not provide a full description of the
source spectrum or power spectral properties since those have been
described in depth for the 2003 outburst in
\citet{2009ApJ...698.1398M} and, for this outburst, in
\citet{2008ATel.1378....1K}. Instead, we use a simple power--law model
to describe the X--ray spectrum in order to derive unabsorbed fluxes
in the 0.5--10 keV band. 

To this end we have extracted spectra from the {\it RXTE}'s
Proportional Counter Array data using {\sl ftools} version 6.6.2.  We
use only data from the Proportional Counter Unit 2 since that was
always operational for these observations. We used the background
model appropriate for bright sources and corrected the spectra for
dead--time, even though the source count rate is such that dead--time
effects are small. We added a systematic error of 0.6 per cent to the
count rate in each spectral bin. Finally, we grouped the spectra such
that there are at least 10 counts per spectral bin. We estimate the
error on the unabsorbed flux that we thus determined to be 10 per
cent.

\begin{table}
\caption{A journal of the {\it RXTE} PCA observations.}
\label{rxtelog}
\begin{center}
\begin{tabular}{lccc}
\hline
Obs ID & Observing  & Start time & Flux  (0.5-10 keV) \\
& date & MJD (UTC) & \flx\ \\
\hline
93427-01-03-03 &2008 Jan.~29  & 54494.19092 & 1.5E-08\\
93427-01-03-04 &2008 Jan.~31  & 54496.74373 & 6.1E-09\\
93427-01-04-01 &2008 Feb.~1   & 54497.85618 & 3.6E-09\\
93427-01-04-00$^a$ &2008 Feb.~2   & 54498.83744 & 8.2E-09\\
93427-01-04-02$^a$ &2008 Feb.~4   & 54500.80062 & 3.8E-09\\
93427-01-04-03$^a$ &2008 Feb.~6   & 54502.82833 & 1.0E-09\\
93427-01-05-00$^a$ &2008 Feb.~8   & 54504.92185 & 9.0E-10\\
\end{tabular}
\end{center}
{\footnotesize$^a$ Data point used in Figure ~\ref{rxcorr}.}\\
\end{table}

\subsection{Results: X--ray decay}

\begin{figure}
  \includegraphics[width=6cm,angle=-90]{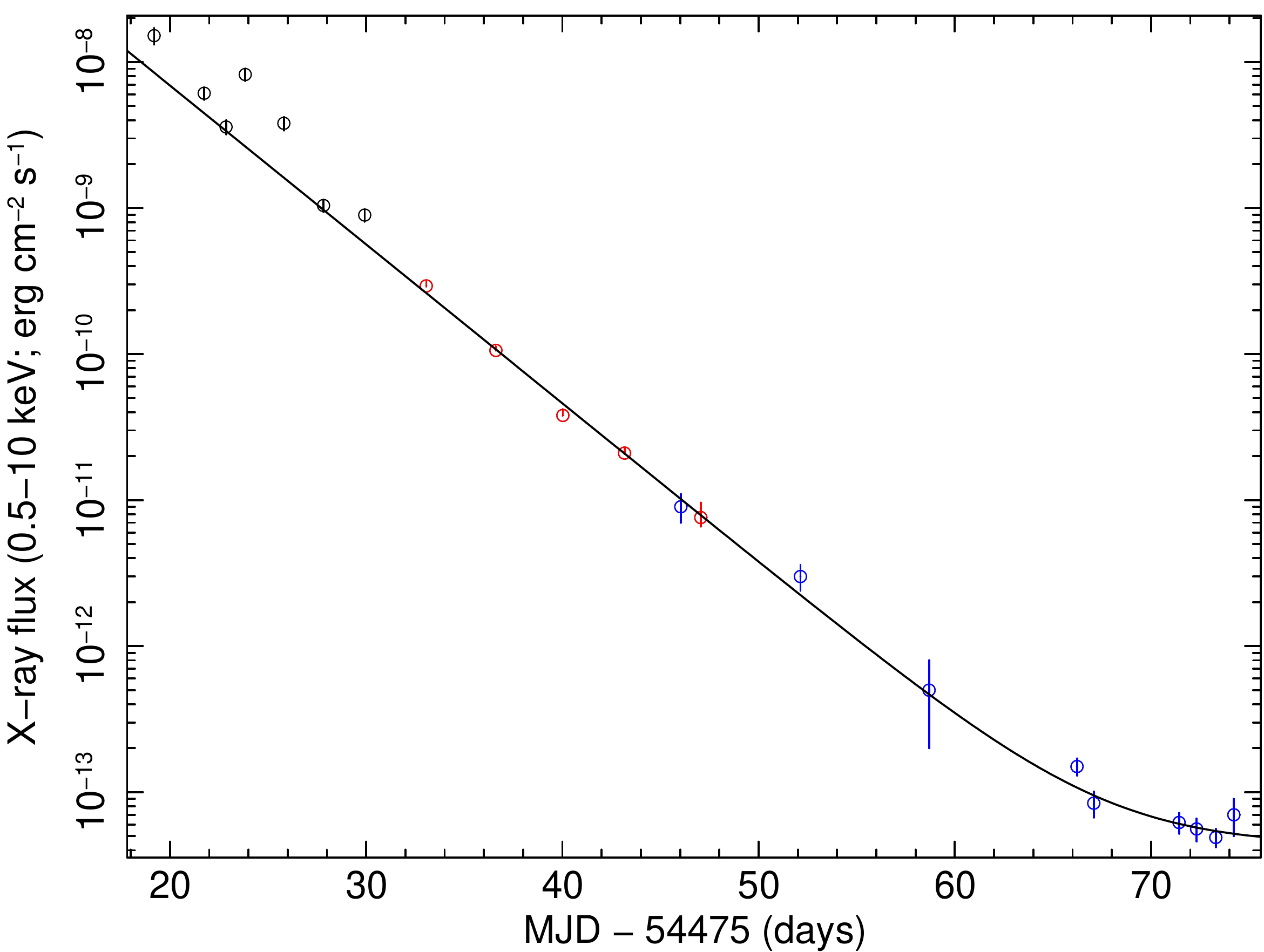}
  \caption{The X--ray light curve of the decay phase of
    the early 2008 outburst of \src\ observed with {\it RXTE} (black
    circles), \swift\ (red data points) and \chan\ (blue data
    points). The best fit of a fit--function consisting of an
    exponential and a constant is overplotted, note that the fit was
    done to the \chan\ and \swift\ data points only and extrapolated
    backwards to the {\it RXTE} data.}\label{xlc}
\end{figure}

In Figure~\ref{xlc} we have plotted the X--ray decay light curve
together with a best--fitting exponential towards the baseline
quiescent flux. The best--fitting parameters are an exponential decay
timescale of $\approx$4 days and a constant of
4.3$\times10^{-14}$\flx. As can be seen from the figure the fit only
globally describes the observations with the $\chi^2$ being very large
(728 for 12 degrees of freedom). Obviously, the X--ray decay light
curve is subject to additional variations beyond that described by the
model (flares and other variability etc). During some of the
\swift\ observations we find strong evidence for such variability. In
contrast, during the \chan\ observations there is no
evidence for such variability. Note that we have not used the {\it
  RXTE} observations in the fit and we have used the \swift\ and
\chan\ flux values determined fixing the ${\rm N_H}$ to 2.3$\times
10^{22}$ cm$^{-1}$ in the figure.

There is a difference in the spectral parameters obtained from the
spectral fits to the \chan\ observations 8985 and 8986 and the
contemporaneous {\it Swift} observations 00031121003 and
00031121005. We attribute this to effects of pile--up in the
\chan\ observations.  The very flat power law and the somewhat low
${\rm N_H}$ are signs of pile--up effects. Furthermore, at the
observed count rates pile--up is indeed expected. The parameters for
the spectral fits to the second \swift\ observation suggest a low
${\rm N_H}$ and a very flat power law index as well. However, the
observed count rate of $<$1 count per second is well below the
threshold of a few hundred counts per second above which the XRT in
windowed timing mode should be suffering from pile--up
(\citealt{2006NCimB.121.1521M}). Close investigation of the spectrum
reveals two potential reasons for the low ${\rm N_H}$ and the flat
power law index. First, there is emission at low energies that is not
present in subsequent observations that are obtained in photon
counting mode. It is unclear whether this emission is real
(cf.~\citealt{2006ApJ...653..525M}) or due to calibration
uncertainties of the windowed timing mode. Nevertheless, even when
discarding all data below 1 keV the best--fitting power law still has
a low index and the ${\rm N_H}$ is low. Second, there seems to be
excess emission around 6.5 keV that can be well described by broad
line emission, however, we lack the signal--to--noise to investigate
the properties of this line in detail. Given that such
(relativistically smeared) emission lines are found in several black
hole candidate sources (cf.~\citealt{2006ApJ...653..525M}) we
attribute the discordant spectral parameters found when the spectrum
is fit with an absorbed power law only, to unresolved line emission
affecting the fit.

\subsection{VLA radio observations and radio decay}

H1743-322 was observed with the Very Large Array (VLA) under
project codes S9208 and AR642. The VLA was in its CnB and C
configurations (for the last two observations) implying a fairly large
beam size ($>$1.5\arcsec).  Observations were carried out in standard
continuum mode with a 50 MHz bandwidth in each of two intermediate
frequency (IF) pairs.  A maximum of 14 retrofitted EVLA antennas were
included in the array.  The primary calibrator was 3C~286, which was
used to set the flux scale according to the coefficients derived at
the VLA by NRAO staff in 1999 as implemented in the 31Dec09 version of
AIPS.  The secondary calibrator was J1744$-$3116 (1 degree from the
target source).  Observations were carried out in fast switching mode
to reduce target--calibrator slew time, using a 3.3~s integration time.
Data calibration and imaging were carried out using standard
procedures within AIPS.  The source flux density was measured by
fitting an elliptical Gaussian to the source in the image plane using
the AIPS task JMFIT.

\begin{figure} \includegraphics[width=6cm,angle=-90]{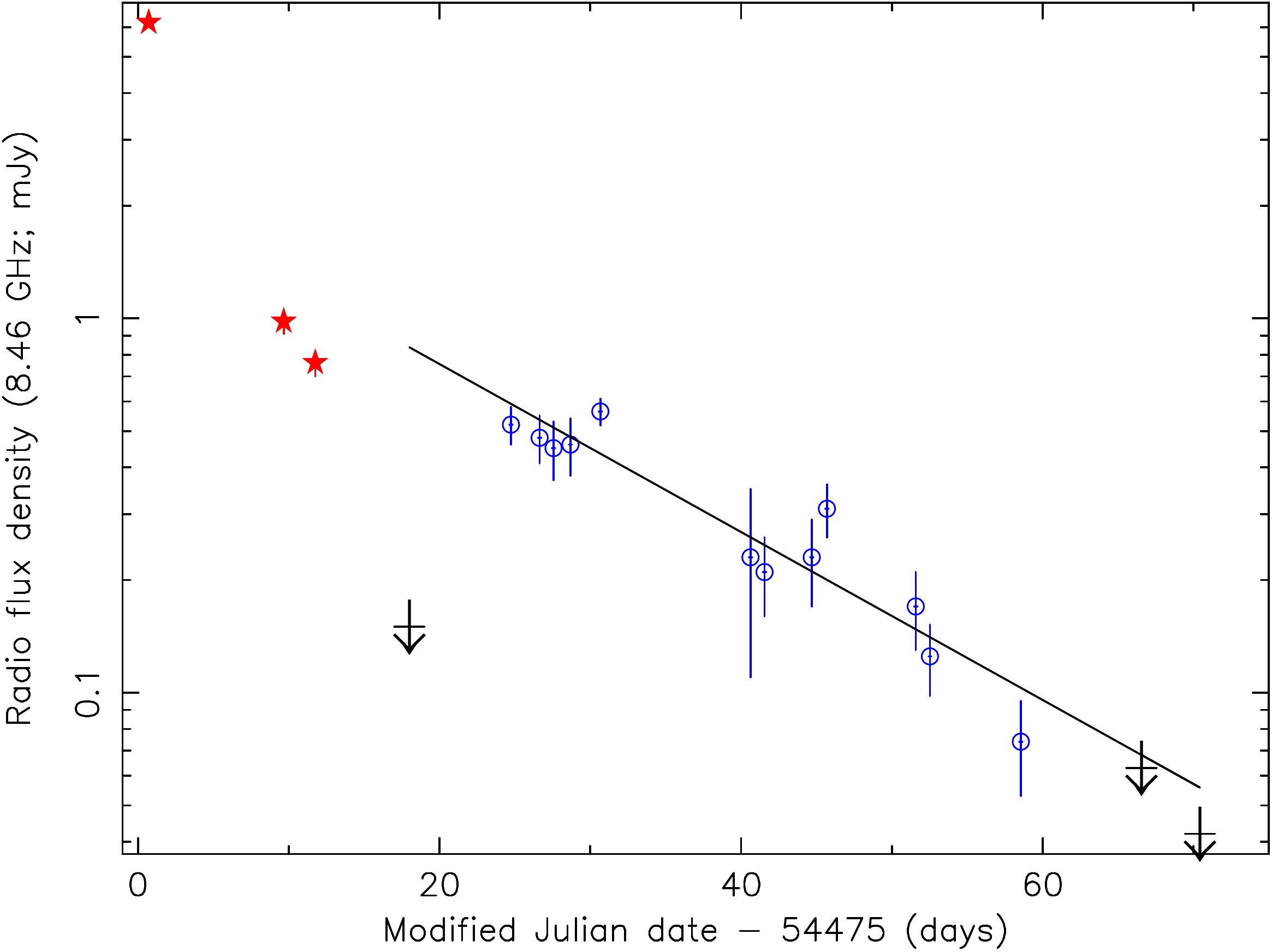}
  \caption{The radio light curve of the decay of the 2008 outburst of
    \src\ observed with the VLA at 8.46 GHz (in addition, the upper
    limit at MJD - 54475 = 18 is an ATCA observation). The first three
    (red) data points indicated by stars are obtained when the source
    was not yet in the low--hard state.  The black arrows indicate the
    3 $\sigma$ upper limit to the radio flux.  The solid line
    represents the best--fitting exponential decay to the low--hard
    state radio light curve. }\label{rlc}
\end{figure}

For a journal of the observations see Table~\ref{vla-log}. In
Figure~\ref{rlc} we plot the observed radio 8.46 GHz decay light
curve. Besides the VLA observations we use one observation obtained
with the Austrialian Telescope Compact Array (ATCA) on Jan.~28, 2008
(MJD~54493) at 8.64 GHz. During this observation the source was not
detected down to a 3 $\sigma$ rms upper limit of 0.15 mJy
(\citealt{2008ATel.1378....1K}; at 8.64 GHz). This stringent ATCA
upper limit shows that the radio emission reactivated close in time to
the transition to the low--hard state.

The solid line in Figure~\ref{rlc} is the best--fitting exponential
decay to the blue open circles, i.e.~the VLA 8.46 GHz detections when
the source was in the low--hard state. The exponential decay timescale
is 19$\pm$2 days. The last radio observation yields a stringent upper
limit of 0.04 mJy, which is below the extrapolation of the
best--fitting exponential. This might indicate that the radio decay
accelerated. The increase in source flux on Feb.~24, 2008 (MJD~54520)
compared with the preceding observations on Feb.~20 and 23, 2008
(MJD~54516, 54519) is atypical since the X--ray flux decayed by a
factor of 15 in the same period. Such behaviour could have been caused
by a radio flare. Note that the radio flare was apparently not
accompanied by an X--ray flare. Alternatively, the X--ray flare
has been missed as the radio observation on Feb.~24, 2008 occurred
1.3 days before a \chan\ observation and 2.5 days after the \swift\
observation nearest in time.

\begin{table*}
\caption{A journal of the radio observations.}
\label{vla-log}
\begin{center}
\begin{tabular}{cccccc}
  \hline
  Calender &  & Frequency & On source time & Flux density & Time between start X-ray\\
  date  & MJD (days) & GHz & (minutes) & (mJy/beam) & and radio obs.~(days)\\
  \hline
2008 Jan.~21     & 54486.740 & 1.4 &3.4& 2.4$\pm$0.26   & XXX\\
2008 Feb.~20     & 54516.544 & 1.4 & 19& 0.6$^b$        & XXX\\
2008 Feb.~23     & 54519.680 & 1.4 & 18& 0.6$^b$        & XXX\\
2008 Mar.~01     & 54526.579 & 1.4 & 18& 0.5$^b$        & XXX\\
  \hline
2008 Jan.~19     & 54484.663 & 4.86&3.4& 1.54$\pm$0.09  & XXX\\
2008 Jan.~21     & 54486.748 & 4.86& 6 & 1.10$\pm$0.07  & XXX\\
2008 Feb.~03     & 54499.737 & 4.86& 17& 0.63$\pm$0.09  & XXX\\
2008 Feb.~06     & 54502.551 & 4.86&2.4& 0.54$\pm$0.12  & XXX\\ 
2008 Feb.~07     & 54503.673 & 4.86&3.4& 0.59$\pm$0.14  & XXX\\
2008 Feb.~09     & 54505.668 & 4.86&93 & 0.52$\pm$0.06  & XXX\\
  \hline
2008 Jan.~10     & 54475.719 & 8.46& 13& 6.16$\pm$0.05  & N.A.\\
2008 Jan.~19     & 54484.669 & 8.46& 6 & 0.98$\pm$0.07  & N.A.\\
2008 Jan.~21     & 54486.760 & 8.46& 10& 0.76$\pm$0.06  & N.A.\\
2008 Feb.~03     & 54499.742 & 8.46& 12& 0.52$\pm$0.06  & 0.9 {\it RXTE}\\
2008 Feb.~05     & 54501.643 & 8.46& 6 & 0.48$\pm$0.07  & 0.8 {\it RXTE}\\
2008 Feb.~06     & 54502.557 & 8.46& 6 & 0.45$\pm$0.08  & 0.3 {\it RXTE}\\
2008 Feb.~07     & 54503.679 & 8.46& 6 & 0.46$\pm$0.08 & 0.8 {\it RXTE} \\
2008 Feb.~09     & 54505.674 & 8.46&17 & 0.56$\pm$0.05 & 0.8 {\it RXTE}\\
2008 Feb.~19     & 54515.625 & 8.46&17 & 0.23$\pm$0.12 & 0.6 \swift \\
2008 Feb.~20     & 54516.552 & 8.46&18 & 0.21$\pm$0.05 & 1.5 \chan \\
2008 Feb.~23     & 54519.689 & 8.46&18 & 0.23$\pm$0.06 & 1.3 \chan \\
2008 Feb.~24     & 54520.694 & 8.46&16 & 0.31$\pm$0.05 & 1.3 \swift \\
2008 Mar.~01     & 54526.587 & 8.46&18 & 0.17$\pm$0.04 & 0.4 \chan \\
2008 Mar.~02     & 54527.529 & 8.46&51 & 0.13$\pm$0.03 & 0.4 \chan \\
2008 Mar.~08     & 54533.556 & 8.46&93 & 0.07$\pm$0.02 & 0.1 \chan \\
2008 Mar.~16     & 54541.550 & 8.46&124& 0.06$^b$ & 0.3 \chan \\
2008 Mar.~20     & 54545.432 & 8.46&355& 0.04$^b$ & 1.0 \chan \\
\hline
2008 Jan.~10     & 54475.720 &22.46&7.5& 3.01$\pm$0.21 & XXX\\

\end{tabular}
\end{center}
{\footnotesize$^a$ VLA 8.46 GHz observation reported in \citet{2008ATel.1384....1R}.}\\
{\footnotesize$^b$ Three $\sigma$ rms limit.}
\end{table*}

\subsection{Radio spectral index}

On several nights near--simultaneous radio data has been obtained at
different frequencies. We use this data to assess the radio spectral
index. Initially, the spectral index is negative, implying optically
thin radio emission. E.g.~the 1.4, 4.86 to 8.46 GHz spectral index on
MJD~54486 is -0.64$\pm$0.07.

After the transition to the low--hard state the spectral index is
consistent with being 0. To obtain a more accurate measurement of the
radio spectral index, we have averaged three short 4.86 GHz and 5 8.46
GHz observations close in time (in the range MJD~54499--54505) where
the radio flux was consistent with being constant. The 4.86--8.46 GHz
radio spectral index for these averages is 0.03$\pm$0.18. This is
consistent with optically thick emission as has been observed in
low--hard radio spectra.

However, the radio spectral index does not stay close to 0 during the
decay to quiescence. We have averaged the 1.4 GHz and 8.46 GHz data in
the MJD range 54516--54526. This data coincides with the radio flare
observable at 8.46 GHz. Combining the three 1.4 GHz observations does
provide a detection at 0.431$\pm$0.097 mJy. At 8.46 GHz the average
flux is 0.175$\pm$0.028 mJy. The 1.4--8.46 GHz radio spectral index is
-0.50$\pm$0.15. This implies optically thin radio emission late in the
low--hard state.

\subsection{Radio -- X--ray correlation}

\begin{figure}
  \includegraphics[angle=-90,width=8cm,clip]{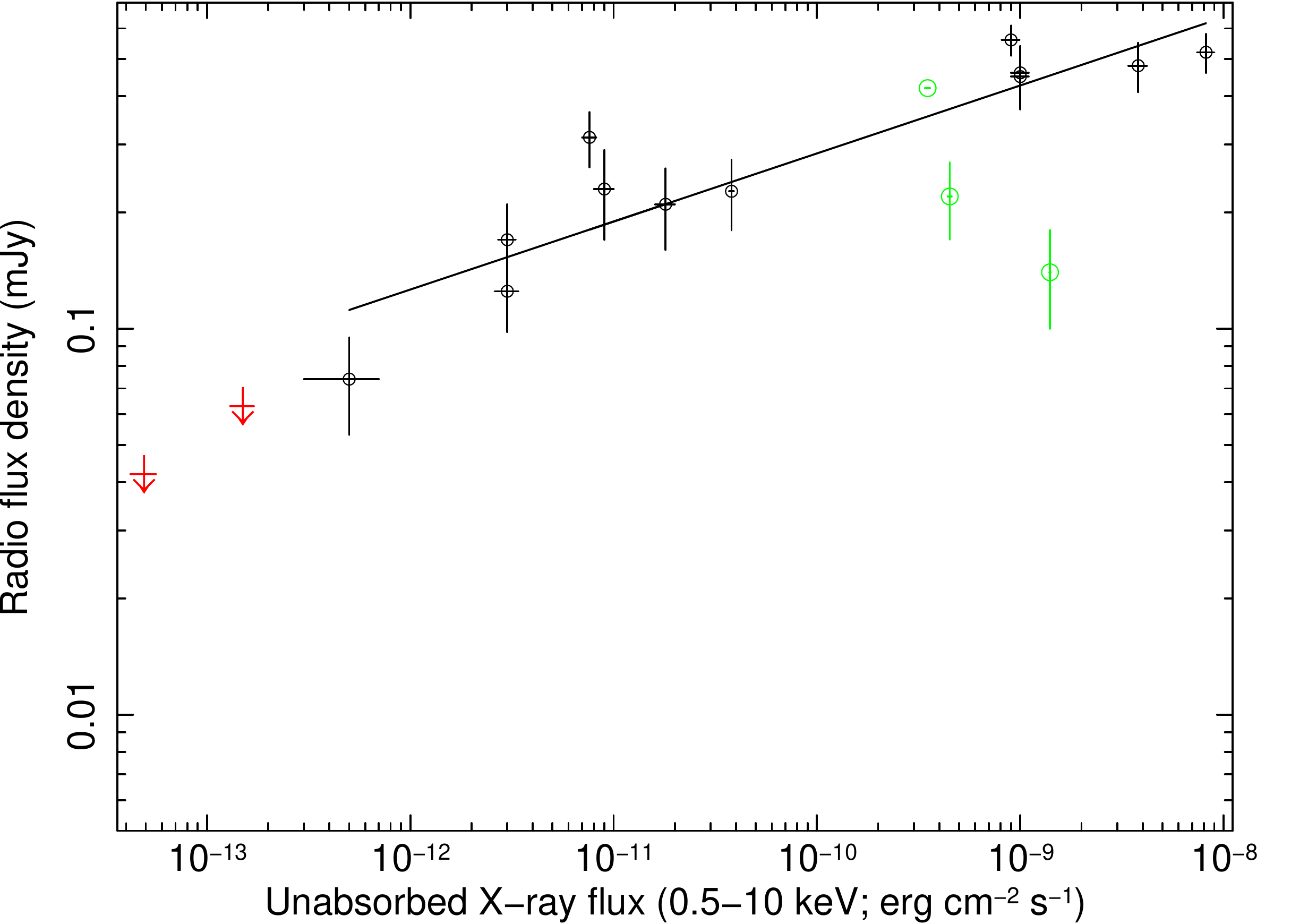}
  \caption{Results on the observed radio and X--ray fluxes of the 2008
    campaign on \src. The observed powerlaw correlation (solid line)
    has a best--fitting index of 0.18$\pm$0.01, much less steep than
    the $L_R \propto L_X^{0.7}$ relation found in several other
    sources.  The correlation does seem to steepen towards the end of
    the outburst. The green circles are the near--simultaneous radio
    and X--ray observations presented by \citet{2009ApJ...698.1398M}
    obtained during the decay after the 2003 outburst when \src\ was
    in the low--hard state.  }
  \label{rxcorr} \end{figure}

In Figure~\ref{rxcorr} we plot the observed correlation between the
X--ray and 8.46 GHz radio fluxes for \src\ using the X--ray
observations closest in time to the radio observations (see footnotes
to the journals of the X--ray observations and Table~\ref{vla-log}).
The best--fitting power law with index 0.18$\pm$0.01 (1 $\sigma$) is
overplotted. The index is less steep than the index of
$\alpha=\approx$0.7 as found for several sources before (S$_\nu
\propto \nu^\alpha; $\citealt{2003MNRAS.344...60G};
\citealt{2003A&A...400.1007C}).

There are several effects that could distort the picture.  First, the
radio and X--ray data are not strictly simultaneous (see
Table~\ref{vla-log}). There is typically less than 1 day between the
start of the radio and X--ray observations, however, in one case the
difference in start times is as much as 1.5 days. This together with the
flares apparent in the radio light curve as well as in the {\it RXTE}
and \swift\ X--ray observations could result in a higher flux in
either radio or X--ray which might confuse the apparent correlation.
Short duration radio flares have been found in e.g.~V404~Cyg
(\citealt{2008MNRAS.388.1751M}). Last, since the spatial resolution of
our VLA observations is several arcseconds, the observed radio flux
might include emission from a ballistic jet--ejection event earlier in
the outburst. Although the ATCA non--detection shows that the initial
radio emission probably related to jet--ejection events occurring
before the transition to the low--hard state faded away, shocks
further down the flow can lead to rebrightenings in radio (and X--ray)
at a later stage. The fact that the radio spectral index during the
epoch related to the radio flare at MJD~54520 is showing that the
radio emission is optically thin strongly argues in favour of this
scenario.

Given the low spatial resolution of our radio observations this
emission can significantly contribute to the radio flux density during
the subsequent measurements. XTE~J1550--564 dramatically demonstrates
this, with radio (and X--ray) emission from the state transition
ejecta still brightening years after the outburst
(\citealt{2002Sci...298..196C}).  The part of the radio decay of \src\
that our observations sample with detections lasts about 33 days. In
the most extreme case of a jet/flow speed of c, in 33 days a distance
of 0.028 pc can be travelled. At a source distance of 7.5 kpc this
corresponds to 0\farcs78 for a jet in the plane of the sky. Such an
angular extent would still be unresolved by the VLA in the CnB and
certainly C configuration.  This indicates that similar to the case of
XTE~J1550--564, shocks in the jet flow could indeed be responsible for
the enhanced radio emission. Since such an angular extend could
potentially be detected by \chan\ we averaged the four \chan\
observations obtained after March 20, 2008. We registered the four
observations to a common astrometric frame using the position of the
brightest X--ray source (besides \src) detected in the observations.
After merging the registered frames, we determined the position of the
X--ray source associated with \src\ in the same way as described
above. We find that the position of \src\ in the average of the last 4
observations in 2008 is consistent with the best known radio position
found by \citet{2004ATel..314....1R}, therefore, the shocked radio
emission is either not responsible for X--ray emission or it has
travelled significantly less than the 0\farcs78 calculated above.

In order for radio emission from shocks downstream to be the cause of
the flattening of the index of the radio -- X--ray correlation, the
radio emission should not be accompanied by a similar amount of X--ray
emission as during the processes responsible for the index of
$\approx$0.7 during the normal low--hard state decay.

In order to compare the normalisation of this correlation with that
found in other sources we need to know the distance of \src\ (see
\citealt{2004MNRAS.inpress}).  As mentioned in the introduction we use
a distance of 7.5 kpc for \src. Comparing the normalisation of the
radio -- X--ray correlation for a distance of 7.5 kpc with that of
other sources we find that \src\ lies above the area traced out by
GX~339--4 and V~404 Cyg (\citealt{2003MNRAS.344...60G}). Even lowering
the distance by a factor 2 one still finds the source above GX~339--4
and V~404 Cyg. Finally, our fluxes and luminosities are given in the
0.5-10 keV band whereas the correlation was plotted in the 2--11 keV
band by \citet{2003MNRAS.344...60G}. The influence of the different
energy bands on the slope of the radio -- X--ray correlation is
minimal since the X--ray spectrum does not change significantly during
the decay. The normalisation is influenced by the difference in energy
bands, though. For comparison we also computed the 2--11 keV X--ray
fluxes of our X--ray observations and we find that the 2--11 keV
values are 80$\pm$5 per cent of the values at 0.5--10 keV. Thus, our
0.5--10 keV luminosities are slightly higher than the 2--11 keV
luminosities bringing \src\ artificially closer to the area traced by
GX~339--4 and V~404 Cyg. The difference in normalisation would be
larger if we had used the exact same energy range.

\section{Discussion}

Using \chan, \swift\ and {\it RXTE} X--ray observations with
(near--)simultaneous radio VLA observations we have observed the decay
towards quiescence of the black hole candidate X--ray binary \src\
while the source was in the low--hard state. The overall shape of the
X--ray decay light curve can be described with an exponential with an
e--folding timescale of $\approx$4 days that levels off towards
quiescence. X--ray flares are superposed on the exponential decay
light curve. During several of the \swift\ observations flares were
observed as well. The unabsorbed 0.5--10 keV quiescent flux is
4.3$\times 10^{-14}$\flx. For an assumed source distance of 7.5 kpc
this implies a quiescent 0.5-10 keV X--ray luminosity of $\approx
3\times 10^{32}$\lum.  The source luminosity in quiescence after the
outburst ending early 2008 is consistent with being the same as that
after the 2004 outburst (\citealt{2005ApJ...632..504C}).

The quiescent X--ray luminosity of $3\times10^{32}$ \lum\ of \src\
implies an orbital period longward of $\approx$10 hours assuming \src\
follows the general trend between orbital period and quiescent X--ray
luminosity reported in \citet{2001ApJ...553L..47G}.  Such an orbital
period suggests that the mass donor star in \src\ has evolved off the
main sequence.  Alternatively, the mass of the black hole in \src\
could be substantially larger than that in the sources defining the
relation between orbital period and quiescent X--ray luminosity.

The low--hard state radio light curve shows evidence for radio flares
superposed on an exponential decay with an e--folding timescale of
$\approx 19$ days.  When plotted against the X--ray light curve the
radio -- X--ray correlation index is lower ($\alpha=0.18\pm0.01$) than
found before ($\alpha\approx 0.7$; \citealt{2003A&A...400.1007C};
\citealt{2003MNRAS.344...60G}).  The radio flares could be responsible
for extended jet emission that due to the relatively large VLA
beamsize in our observations using C and CnB configurations was not
resolved. The 1.4--8.46 GHz radio spectral index during the MJD~54520
flare is -0.50$\pm$0.15 indicating that the radio emission is probably
optically thin.

Relativistic arcsecond scale jets have not been found before for
sources in the low--hard state (cf.~\citealt{2006csxs.book..381F};
\citealt{2009MNRAS.396.1370F}).  Therefore, it seems more likely that
jet ejections earlier in the outburst of \src\ caused shocks when
travelling along the jet.  We note that similar events could be behind
the apparent scatter and variability in the observed radio -- X--ray
correlation in other sources (cf.~\citealt{2008MNRAS.389.1697C}). The
resulting index then depends on the amount and energy in the shocks
caused by earlier ejection events. These will change from source to
source and from outburst to outburst. For instance, plotting the radio
and X--ray detections in the low--hard state during the decay
\citet{2009ApJ...698.1398M} we find that two of the three 2003 data
points fall approximately a factor 2 to 3 below the curve in
Figure~\ref{rxcorr}.

Increases in the radio flux density in the low--hard state by factors
of 3--10 have been observed in the short--timescale radio flares in
the quiescent state of V~404 Cyg (\citealt{2008MNRAS.388.1751M}).
Given that the radio and X--ray data that we presented here are not
exactly simultaneous, any hour--long radio flare would have
no corresponding flare in the X--ray band. Furthermore, even if the
observations had been strictly simultaneous, the short integration
times of our initial radio observations implies that an X--ray flare
could have been missed since such an X--ray flare would take tens of
minutes or longer to propagate downstream to the point in the compact
jet where the optical depth is $\approx$1 at 8.46~GHz. Nevertheless,
in order to produce a slope of $\approx 0.2$ the radio observations
must have been close to the peak of flares, which seems unlikely.  We
conclude that optically thin radio emission from shocks caused by
jet ejections earlier in the outburst influence the late time low--hard
state radio emission.

Previous studies of black hole (candidate) sources decaying via the
hard state to quiescence suggests that the power--law index of the
X--ray spectrum in the quiescent state is different from that in the
low--hard state.  \citet{1994PASJ...46..375E} observed GS~1124--68
with {\it Ginga} and found a power-law index that remained essentially
constant at $\sim$1.6 during the time the source was decaying in the
hard state, except at the lowest observed luminosity of $\sim5.5\times
10^{-5}$\,\ledd\ (for $M_{BH}$=7$M_\odot$) when it was found to be
1.84$\pm$0.04.  \citet{2003A&A...400.1007C} have reported a power-law
spectral index of $\sim$2.2 around $5.3\times 10^{-6} (d/6\,{\rm
  kpc})^2$\,\ledd\ for GX~339--4 (for $M_{BH}$=5.8$M_\odot$).
\citet{2004ApJ...601..439T} followed XTE~J1650--500 down to
$\sim1.4\times 10^{-5} (d/6\,{\rm kpc})^2$\,\ledd\ (for
$M_{BH}$=8.5$M_\odot$) with {\it Chandra}, and also found evidence for
spectral softening. Finally, \citet{2008MNRAS.389.1697C} show that the
power--law spectral index is softer in V404~Cyg in quiescence compared
to that in the brighter low--hard state.

\begin{figure}
  \includegraphics[width=6cm,angle=-90]{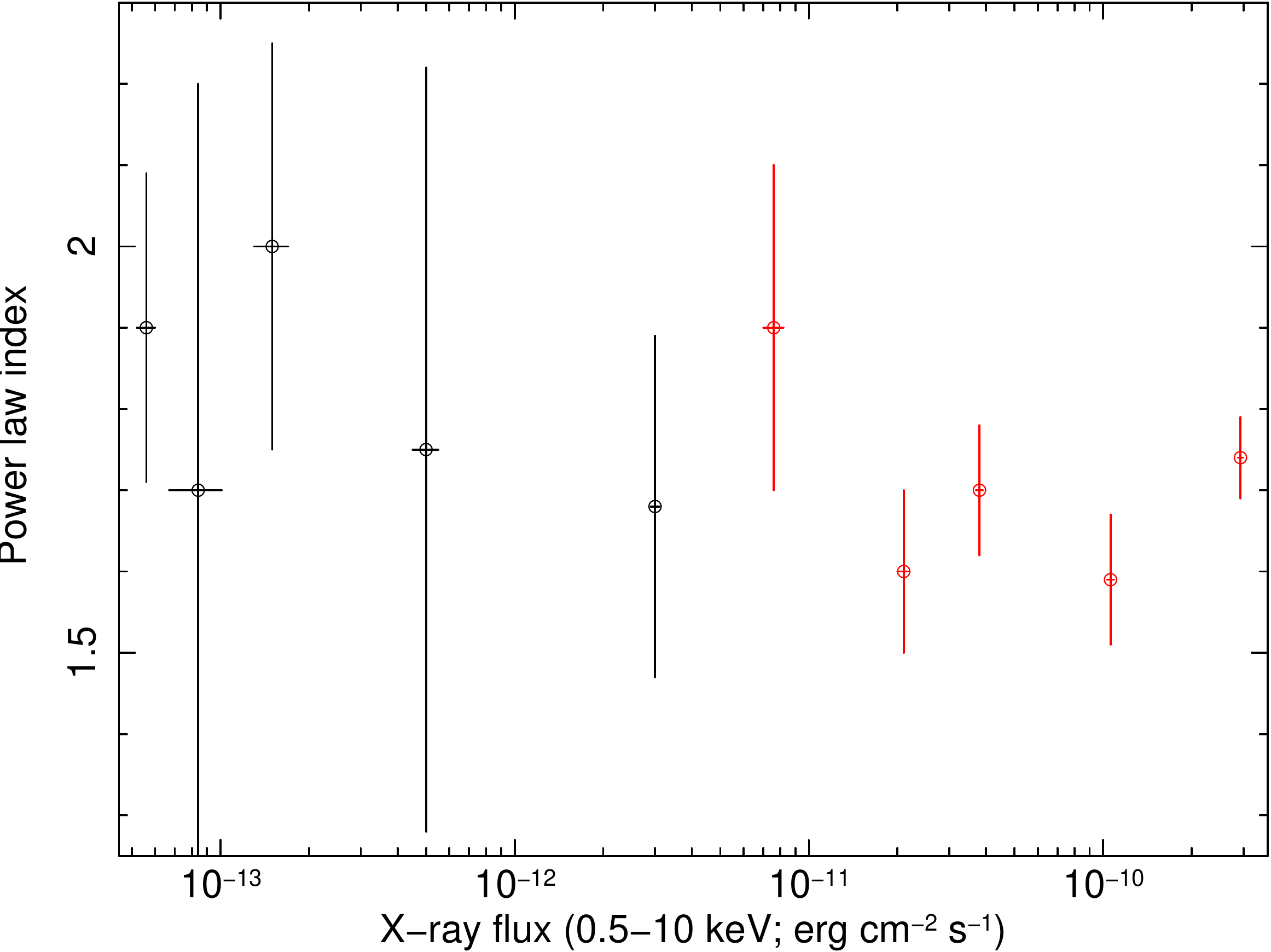}
  \caption{The X--ray power--law index observed with \swift\, and
    \chan\, of \src\, as a function of source flux during the last
    part of the outburst of \src.  The red data points are \swift\
    measurements and the black data points are \chan\ measurements
    (the lowest--flux data point includes \chan\ data presented in
    Corbel, Tomsick \& Kaaret 2006).}\label{plindex} \end{figure}

In Figure~\ref{plindex} we have plotted the power law index from the
\swift\ and \chan\ spectral modelling on \src\ during the decay
towards quiescence and in quiescence.  The data point at the lowest
X--ray flux is determined by combining the quiescent data obtained
right after the outburst of early 2008 as well as the quiescence
observations used by \citet{2005ApJ...632..504C}. There is no evidence
for a softening of the power--law index towards and in quiescence,
although for this source the relatively high neutral hydrogen column
density limits the accuracy of the power--law determination.  A fit of
a constant power--law spectral index gives a best--fitting index of
1.704$\pm$0.008 with a $\chi^2=7$ for 9 degrees of freedom.

We have compared the X--ray flux decay rate of \src\ with that
observed during the last part of the outburst decay from the black
hole candidate XTE~J1908+094 and we find that the fit function
describing the decay of \src\ is a good approximation to the last
phase of approximately three weeks presented in Jonker et al.~(2004)
for XTE~J1908+094. Note however, that due to a reduced sampling that
source was only observed twice during those three weeks. Nevertheless,
it is interesting to see that the sources have a similar decay rate.
If other sources also follow the same decay rate in X--rays this could
provide constraints on the accretion disc model, such as the accretion
disc to ADAF evaporation.

\section*{Acknowledgments} \noindent PGJ acknowledges support from a 
VIDI grant from the Netherlands Organisation for Scientific
Research. MAPT acknowledges support from NASA grant GO8-9042A. DS
acknowledges an STFC Advanced FellowshipThe National Radio Astronomy
Observatory is a facility of the National Science Foundation operated
under cooperative agreement by Associated Universities, Inc.

\end{document}